# An Ice-Structuring Mechanism for Zirconium Acetate

Sylvain Deville[*], Céline Viazzi[†], Christian Guizard

Laboratoire de Synthèse et Fonctionnalisation des Céramiques, UMR3080 CNRS/Saint-Gobain, Cavaillon, France

**Abstract**

The control of ice nucleation and growth is critical in many natural and engineering situations. Yet, very few compounds are able to interact directly with the surface of ice crystals. Ice-structuring proteins, found in certain fishes, plants and insects, bind to the surface of ice, thereby controlling their growth. We recently revealed the ice-structuring properties of zirconium acetate which are similar to those of ice-structuring proteins. Being a salt, and therefore different from the proteins having ice-structuring properties, its ice-structuring mechanism remains unelucidated. Here we investigate this ice-structuring mechanism through the role of the concentration of zirconium acetate and of the ice crystal growth velocity. We then explore other compounds presenting similar functional groups (acetate, hydroxyl, or carboxylic groups). Based on these results, we propose that zirconium acetate adopts a hydroxy-bridged polymer structure which can bind to the surface of the ice crystals through hydrogen bonding, thereby slowing down ice crystal growth.

Introduction
Very few compounds are able to interact directly with ice, due to the very subtle structural differences between water molecules and the ice surface. Yet, such interaction is often critical in many natural or engineering phenomena, to prevent, restrict or control ice growth. Compounds known in biology as antifreeze proteins are currently found in certain living organisms (fishes, plants, insects) where their interaction with the ice crystals prevents fatal freezing-induced damage [1]. This ensures survival at low temperature and provides an evolutionary advantage. Understanding the exact mechanism by which they prevent or control the growth of ice crystals is still a major challenge in biology.

Classical antifreeze proteins exhibit three characteristic behaviors, (1) thermal hysteresis (from a non-colligative reduction in the freezing point, thus the name antifreeze), (2) ice recrystallization inhibition, and (3) ice structuring (change in the morphology of ice crystals). This has led to alternative names such as thermal hysteresis proteins and ice structuring proteins (ISPs)[2,3], to identify groups of proteins that may not exhibit all three of these properties.

The current explanation is that ISPs bind to a specific family of crystals planes thereby restricting their growth and resulting in facetted ice crystal growth. An anchored clathrate mechanism has been recently proposed [4] for the binding.

The control of ice crystal growth is also of interest for problems as diverse as texture control in frozen food and deserts [5], cryopreservation of cells and tissues, cryosurgery adjuvants, and materials engineering [6–8]. Inspired by ISPs, ice-structuring compounds have been developed as a cheaper and more convenient alternative. Among the structural requirements are the presence of hydrophobic and hydrophilic regions in the structure, long range order, and the absence of charged functional groups [9]. Such compounds are currently commercially available and used, although they are still less efficient than ISPs. All ISPs discovered to date are long organic macromolecules, often with an amphiphilic structures.

We recently revealed how a radically different compound, zirconium acetate (ZRA), exhibits ice-structuring properties similar to that of ISPs [10]. ZRA is a metallic salt, and salts were not known to exhibit such properties. The ice-structuring properties of ZRA suggest that it somehow interacts directly with the ice crystals at the atomic level. The presence of additional compounds (binder, dispersant) or particles is not required to observe the ice-structuring properties.

In materials science, the growth morphologies of ice crystals can be used to control the pore morphologies in ice-templated materials, where the pores are a replica of the ice crystals [11]. Generally, the growth behavior is dependent on a number of process parameters, such as the suspension viscosity and pH [12], the particle concentration, the zeta potential [13,14] and the growth velocity of the ice crystals [15–19]. Additives can be incorporated to modify these characteristics, although their influence is mostly unpredictable and dependent, to some extent, on the suspension and ceramic powder properties [20].

We take advantage of the ice templating effect to investigate the ice-structuring mechanism by which ZRA controls ice crystal growth. We explore the influence of ZRA concentration and ice crystal growth velocity. We then assess whether other compounds presenting similar functional groups –acetate, hydroxyl and carboxylic groups– exhibit the same properties. Based on these results, we propose an ice-structuring mechanism similar to that of ISPs.

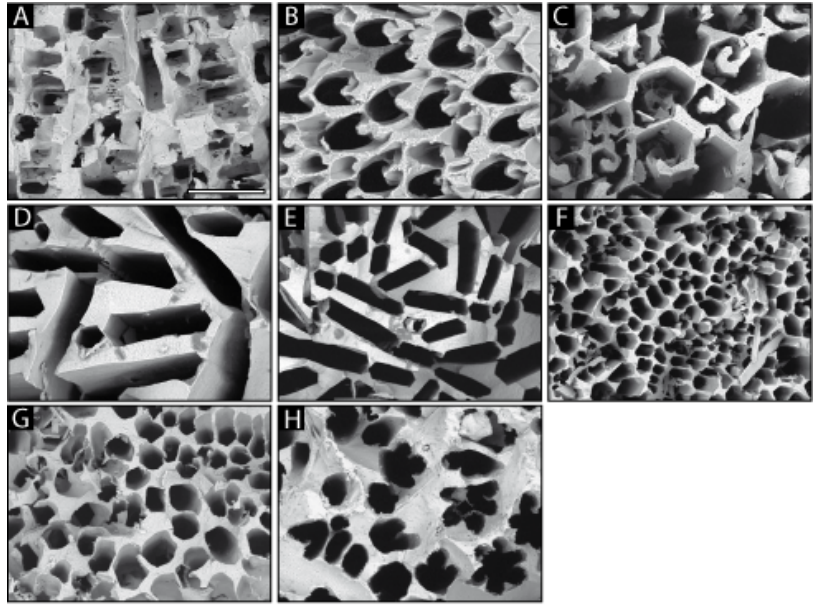

Figure 1. Influence of ZRA concentration on the pores morphology. The black regions (pores) are replicas of the ice structure. Scale bar 50 μm. The respective concentrations (in g.l$^{-1}$ of Zr) are a: 0, b: 4.5 g.l$^{-1}$, c: 9 g.l$^{-1}$, d: 18 g.l$^{-1}$, e : 45 g.l$^{-1}$, f : 180 g.l$^{-1}$, g : 271 g.l$^{-1}$, h : 453 g.l$^{-1}$. Cooling rate 5°C.min$^{-1}$.

**Experimental methods**

The following compounds were bought from Sigma-Aldrich (St. Louis, MO, USA): zirconium acetate, zirconium acetate hydroxide, Pluronic F-127, pentaerythritol, D-sorbitol, xylitol, dextrin, beta-lactose, D-quinic acid, myo-inositol, acetic acid and propanoic acid. Yttrium and barium acetate were bought from Alfa-Aesar (Ward Hill, MA, USA). The pH of the suspension was adjusted using addition of two different acids (HCl or HNO$_3$) and two different bases (NaOH or KOH), to rule out the possible influence of the ionic species. The suspension preparation, freezing and freeze-drying methods, briefly summarized here, are identical to that of reference 10.

The suspensions are prepared by mixing the ceramic powder or the polymer in aqueous suspension, the zirconium acetate and the binder in distilled water. Experiments are carried out with yttria-stabilized zirconia (TZ8Y, Tosoh, Japan) as a powder. The zirconium acetate tested, is commercially available (Sigma-Aldrich, St. Louis, MO, USA). The amount of zirconium acetate in the suspension corresponds to a concentration of zirconium provided by the zirconium acetate (g/L). Zirconium acetate is first mixed with distilled water. The solid loading of the suspension is 55wt%. The binder is then added, preferably after having been dissolved in water, and the ceramic powder is finally added. The suspension is then ball-milled for 10 hours.

The suspension is poured into a PTFE mould and cooled from the bottom, using a liquid-nitrogen cooled copper rod. The cooling rates are adjusted through a thermocouple and a ring heater placed around the copper rod. Details of the experimental setup can be found in previous papers, such as reference 14,17. This setup provides a fine control of the ice growth velocity, along the temperature gradient direction. Once freezing is completed, the samples are freeze-dried for at least 48 hrs in a commercial freeze-dryer to ensure a complete removal of the ice crystals. A binder removal step is performed with the following cycle: temperature rise at a rate of 600°C/h up to 500°C, steady stage of 1 hour at 500°C, temperature decrease to room temperature. Ceramic samples are densified by a high temperature sintering treatment, with a temperature rise at a rate of 600°C/h up to 1350°C, steady stage of 3 hours at 1350°C, temperature decrease at a rate of 600°C/h to room temperature. Samples are cut, perpendicular to the direction of the temperature gradient, using a low-speed diamond saw, and cleaned in an ultrasound bath to remove the debris. SEM observations were performed using either a TM1000 from Hitachi or a Nova NanoSEM 230 from FEI.

## Results and discussion

In the following sections, the reader should keep in mind that the pores are a replica of the ice crystals. The templating of ice crystals is thus used to investigate the morphology and characteristics of the ice crystals.

This process has already proved to be extremely consistent and robust. Provided that the formulation of the suspension is the same, using the same cooling setup and cooling rate yield structures with the same morphology. No variations of morphology are found across samples obtained in the same conditions. This consistency is obtained because of the equilibrium conditions obtained during the unidirectional solidification[21]. After the initial nucleation and growth stages (transient regime)[22], a stable equilibrium regime is obtained. This approach is very different to the ones usually used in the investigation of ice-structuring properties of ISPs, where nucleation and growth of ice is carefully controlled and isolated, single crystals are observed.

## Influence of zirconium acetate concentration

Pore morphology is extremely dependent on the concentration of ZRA in solution. Without ZRA, the pores exhibit a convoluted structure, highly dendritic and with a significant surface roughness (fig. 1a). Such dendritic structure is similar to that already reported in the literature. Above a concentration threshold (18 g.l$^{-1}$)(fig. 1d), faceted pores are obtained, as reported previously10. Faceting occurs over a large range of concentration above the threshold and increasing the concentration has little effect over the porous structure (fig. 1d, 1e and 1f). Faceting progressively degrades when the ZRA concentration is very high, greater than 200 g.l$^{-1}$ (fig. 1g and 1h).

A low concentration of ZRA (9 g.l$^{-1}$), below the concentration threshold required for faceting (18 g.l$^{-1}$), impacts the pores morphology and provides additional clues on the ice crystal structuring mechanism which are discussed in the next section. The pore structure is dramatically different from that obtained without ZRA. Just below the threshold (fig. 1c), the pores exhibit a dual structure: some faceting can be observed on the outer part of the pores, while the inside reveals a spiral-like structure.

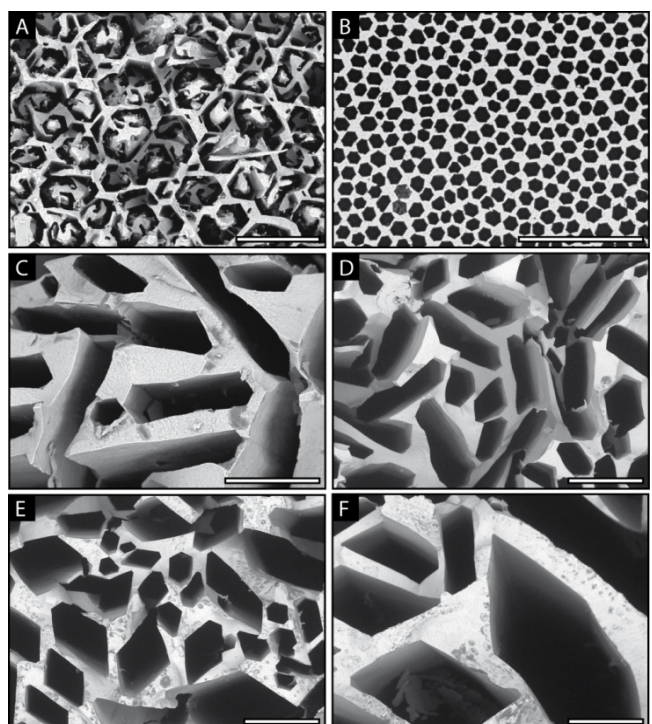

Figure 2. Influence of ice crystal growth velocity (a) >20°C min$^{-1}$, (b) 15°C min$^{-1}$ (c) 5°C min$^{-1}$ (d) 2°C min$^{-1}$ (e) 0.5°C min$^{-1}$ (f) 0.2°C min$^{-1}$. Scale bar (a-c) 50 µm (d-f) 150 µm. Concentration: 18 g.l$^{-1}$ ZRA.

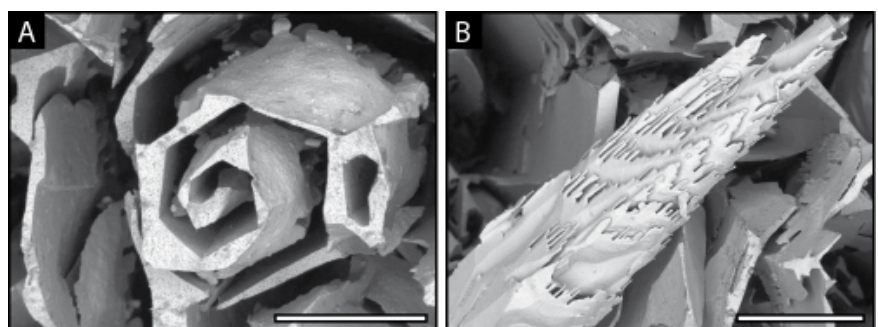

Figure 3. Structural features below the faceting threshold, for very slow crystal growth rate. The suspension was maintained at -5°C to obtain a very slow growth rate. Features typical of a spiral growth mode, seen from top (a) and from aside (b). Scale bars: (a) 50 µm, (b) 250 µm. Concentration: 18 g.l$^{-1}$ ZRA.

## Influence of ice crystal growth velocity

The ice growth velocity, resulting from the imposed cooling rate and temperature gradient, has a dramatic effect on the pores, and yields different morphologies. Faceting is found for cooling rate lower than 15°C min$^{-1}$. For fast cooling –15°C min$^{-1}$–, honeycomb-like structures are obtained with homogeneous, hexagonal pores of a few micrometers (fig. 2b). Decreasing the cooling rate to 5°C min$^{-1}$ triggers elongated growth of the faceted pores along one direction (fig. 2c). In cross-sections perpendicular to the solidification direction, the thickness of the pores remains essentially constant, while their length increases. Further decrease of the cooling rate to 2°C min$^{-1}$ or below yields faceted microstructures with heterogeneous pore size (fig. 2d-2f). Conversely, increasing the cooling rate to above 20°C min$^{-1}$ yields microstructures with pores exhibiting faceted surfaces on their outer part and spiral features in their inner part (fig. 2a), a behavior similar to that observed when the ZRA concentration is near the faceting threshold. Adjacent and opposed growth spirals, possibly due to pairs of dislocation, can be observed in figure 2a.

## Growth mode

Local observations (fig. 3) reveal details about this spiral structure. Such structures are typical of a spiral growth mode which occurs around a screw dislocation or in the presence of instabilities of the crystal edge relative to the crystal surface which also lead to hopper morphologies[23]. In addition, we observe that the outside of the crystal –hence, the inside of the ceramic spiral– is facetted, while small dendritic features can be observed along the spirals (fig. 3b), a situation that can be found when the driving force is the greatest along the edges or corners of the crystal. This is possibly another indication that the ZRA is modifying the growth driving force, a behavior similar to that of ice-structuring proteins.

The progressive degradation of the faceting at high ZRA concentration (fig. 1g and 1h) can be understood as a modification of the solvent. ZRA is provided in acetic acid solution. For high ZRA concentration, the solvent is not pure water anymore, but a mix of water, acetic acid and ZRA. It is thus possible that new phases are formed in this system, though we have no knowledge of the corresponding phase diagram. The crystals formed could thus be crystals of a phase comprising water and acetic acid, and eventually ZRA in some form. The crystal structure is likely not hexagonal anymore, explaining the loss of the facetted structure. The role of acetic acid and compounds with similar functional groups is further explored later in this paper.

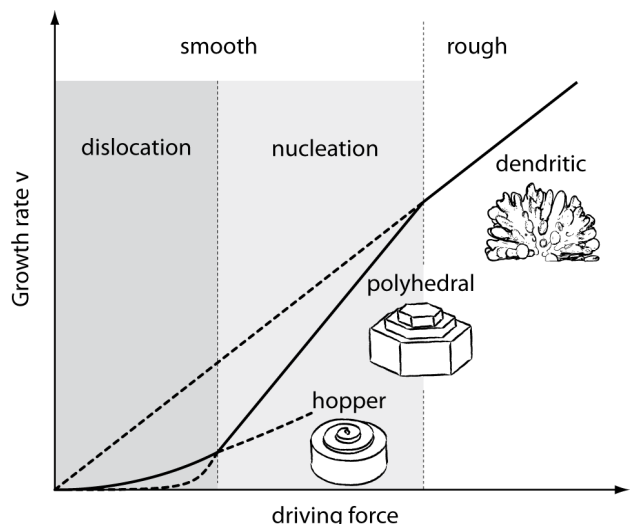

Figure 4. Schematic relationships between crystal growth rate and driving force, and corresponding crystal morphologies for rough (continuous growth) and smooth (hopper and polyhedral) morphologies. After [24].

The evolution of the pore morphology with the ZRA concentration or growth velocity is reminiscent of the usual relationships between crystal growth rate and driving force reflecting an evolution of the interface morphology from rough to smooth (fig. 4) and therefore a change of the growth regime or of the rate-controlling process. Dendritic (figure 1), polyhedral (figures 1d and 1e) and hopper (figure 3) structures are observed.

The concentration of ZRA and the cooling rate thus have a similar effect on the crystal growth regime. Although the relationship between growth rate and driving force (undercooling) is well-known in crystal growth, the mechanism controlling the undercooling in presence of ZRA is still unknown. The evolution from a rough to a smooth interface implies a change of rate-controlling process. Rough interfaces occur when the diffusion of molecules or atoms to the growth surface is the rate-controlling process. For smooth interfaces, the adsorption of atoms or molecules at the surface is the rate-controlling process. Adsorption of zirconium acetate could thus explain the current observation. The functional groups of zirconium acetate should thus play a critical role in the ice-structuring mechanism.

## Assessing the role of functional groups

Zirconium acetate is a compound radically different from the known ice structuring compounds. We investigated the role of the functional groups to assess their influence on the ice structuring mechanism. The ZRA used in this study is provided in an acetic acid solution; three kinds of functional groups are thus present. We tested compounds respectively acetate, hydroxyl and carboxyl groups, these groups being present, respectively, in ZRA and acetic acid (fig. 5). All compounds were tested in a large concentration range, 1–20 wt.% and in the pH range 2-8.

The closest compound to ZRA is probably zirconium propionate, but is insoluble in water and thus could not be tested. Zirconium hydroxyl acetate has also a very similar structure to zirconium acetate, and it does not induce any faceting (fig. 6c). Neither do acetates such as yttrium acetate or barium acetate (fig. 6a and 6b). We investigated the role of carboxyl groups with acetic and propanoic acid. None of these compounds induce any faceting (fig. 6d and 6e).

Some compounds with hydroxyl groups, such as polyols, are known antifreeze compounds. Examples include D-sorbitol, xylitol or pentaerythritol. In addition, we also tested dextrin, be-ta-lactose, D-quinic acid, myo-inositol, and a block copolymer (Pluronic F-127). Faceted structures have not been obtained with any of these compounds (fig. 6f to 6m). All polyols seem to have a similar effect (fig. 6g to 6m), yielding dendritic structures, similar to that obtained with sucrose in previous experiments [20,25]. The block copolymer is apparently the only one exhibiting an ice-structuring effect, the corresponding structures having much smaller pores (fig. 6f), corresponding to much small ice crystals. It might therefore effectively decrease the ice crystal growth rate, though it does not induce any faceting of the ice crystals.

## Influence of poly(vinyl alcohol)

PVA is routinely added in ceramic processing as a binder and was used here in the processing of the samples. The antifreeze properties of non-peptide polymers, including poly(vinyl alcohol) (PVA), poly(ethylene glycol) (PEG), poly(acrylic acid), poly(L-hydroxyproline) or poly(L-histidine), have been investigated [26–30]. They have been found to exhibit some or all of the three antifreeze mechanisms of antifreeze proteins: recrystallization inhibition of ice, thermal hysteresis, and ice crystal morphology control. The role of PVA on the observed ice structuring properties must be carefully considered. PVA is of particular interest, as it is the only non-peptide polymer that has been reported to cause thermal hysteresis. In addition, dynamic ice structuring properties have also been observed, with the growth of faceted crystals [27]. Being known for its antifreeze properties, PVA has naturally been tested in ice templating. Although it has been found to limit the growth of ice crystals, faceted ice-templated structures have not been observed [31–33]. Its ice-structuring properties appear therefore to be less prominent than that of ZRA.

Salts, including acetates, have been reported to enhance the effects of these macromolecules in binding to ice [34], but do not have ice structuring activity on their own. Although ZRA could therefore possibly increase the antifreeze activity of the binder (PVA or PEG) used in ice-templating, ZRA was found to exhibit ice-structuring properties on its own. No differences were observed when the binder was incorporated into the suspension [10]. Thus, the ice-structuring properties of ZRA cannot be explained by the presence of PVA. The antifreeze properties of PVA can contribute to that of ZRA but are not predominant.

## Structure and ice-structuring mechanism of ice structuring compounds

ZRA seems to exhibit strong functional similarities to ice- structuring compounds. To better understand the ice-structuring mechanism on ZRA, we briefly review the current understanding of ISPs. ISPs control the growth of the ice crystals at the atomic level. Some commercially available ice-structuring compounds can prevent or slow down the growth of the ice crystals, used particularly in food engineering[35,36], but their action does not seem to be as specific as ISPs. Although a wide variety of ISPs have been found to date, they all share some similarities such as common moieties and functional groups. These long macromolecules comprise amphiphilic groups. In some cases, alternation of hydrophilic and hydrophobic groups is observed. The conformation of ISPs often matches the structural periodicity of particular crystallographic planes of the ice crystals [37]. A close contact between the surface of the ice crystals and the ISPs is thus possible, enhancing Van der Waals interactions. Various mechanisms have been proposed based on these observations. In particular, it has been proposed that the matching of the protein to the ice surface prevents or slows down the incorporation of water molecules at the surface of the ice crystals [1,38,39] in a manner similar to particles in front of a solidifying interface [40]. Depending on their structure, ISPs can interact with a variety of ice's crystallographic planes. In some cases, growth is favoured along the c–axis, while other compounds yield the opposite behaviour. In all cases, the ISPs form an ice-interacting network effectively slowing down crystal growth.

## Ice-structuring mechanism of zirconium acetate

Comparisons between ISPs and ZRA can be drawn to understand the ice-structuring properties of ZRA. First, they share some similarities. The moieties of ZRA are similar to those of some ISPs due to the presence of amphiphilic groups. The structure and conformation of the compound is nevertheless critical for the ice-structuring mechanism, as the experimental results obtained with the alternative compounds sharing similar functional groups did not reveal ice structuring properties. It is thus clear, although not surprising, that moieties alone without a periodicity of their location along the compound cannot induce ice structuring. Both ZRA and ISPs require a minimum concentration below which no ice structuring properties are observed. A kinetic effect is also observed where ice-structuring properties disappear when the ice crystal growth velocity exceeds a threshold. Faceted growth is observed in both cases, likely implying control at the atomic level. With unidirectional solidification, the c-axis aligns along the temperature gradient 10 similar to the behaviour of some ISPs where growth along the c-axis is favoured.

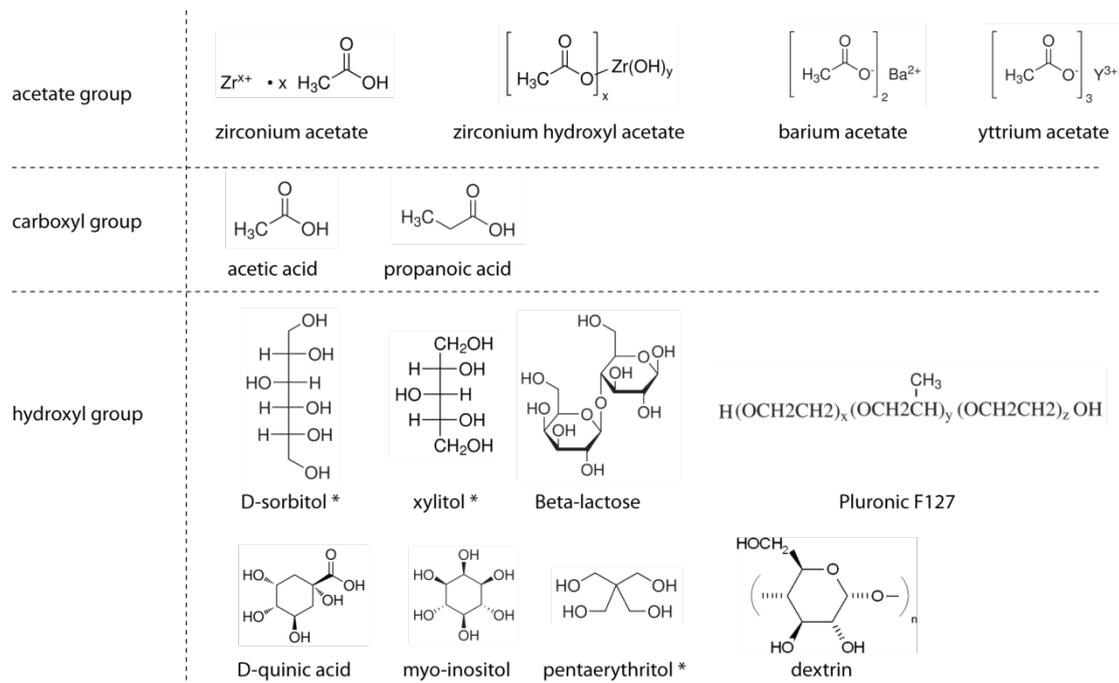

Figure 5. Investigated compounds. The stars indicate compounds known to have some antifreeze activity.

ZRA is used in very different applications. In materials processing, besides its influence on the viscosity of the suspension [41] or its use as a precursor of zirconium [42] in sol-gel or catalysis applications, it is also used as a depletion induced stabilizer of colloidal suspension [43], including carbon nanotubes suspensions [44]. Since ZRA is controlling the growth of ice crystals without any other additives or particles [10], the depletion properties cannot account for its ice-structuring properties. It is also used for waterproofing [45] and flame retardant treatment [46], in particular in paper and textiles applications, for cross-linking between textile substrates and films of water-repellent compounds.

Zirconium acetate solution is believed to contain neutral species. We believe that under the conditions where ZRA exhibit ice-structuring properties, it adopts the structure of a hydroxy-bridged polymer [47,48], that can be anchored to the surface of the ice crystals via hydrogen bonds (figure 7). Our experimental observations during ice templating can thus be explained by the proposed ice-structuring mechanism. A minimum concentration of ZRA is required to ensure a minimal coverage of the ice surface. The disappearance of ice-structuring properties at high crystal growth velocities could be understood by kinetic considerations. A certain time is required for the adsorption to be effective, and thus cannot occur if the crystals are growing too fast.

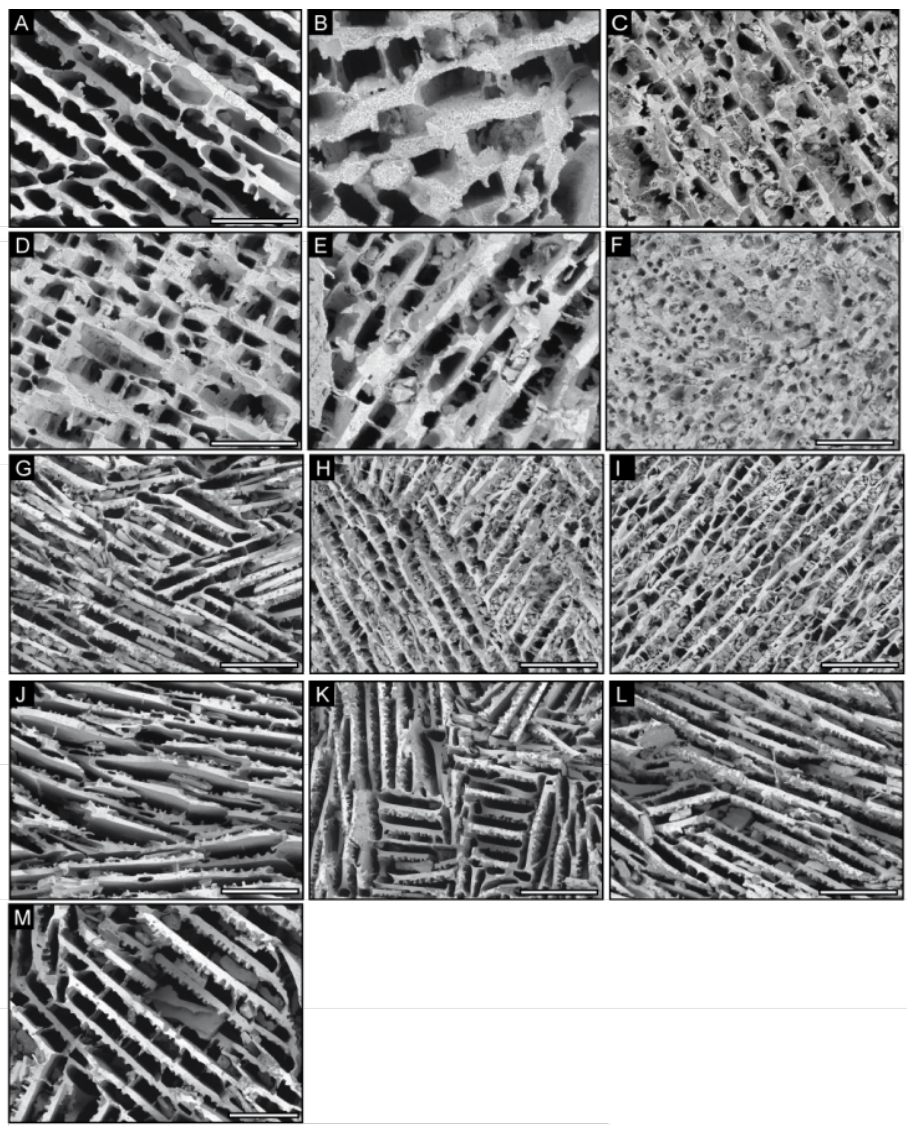

Figure 6. Assessing the role of acetate groups (a) 2 wt.% yttrium acetate (b) 6 wt.% barium acetate (c) 5 wt.% zirconium hydroxyl acetate. (d) 1.5 wt.% acetic acid, (e) 1.8 wt.% propanoic acid, (f) Pluronic F-127; 5 wt.% of (g) D-sorbitol (h) xylitol (i) pentaerythritol (j) dextrin (k) beta-lactose (l) D-quinic acid (m) myo-inositol . Scale bar a-f 50 μm, (g) 250 μm (h-j) 100 μm (k-l) 250 μm (m) 150 μm.

Two different binding mechanisms can be considered for the hydroxy-bridged polymer structures, depending on whether hydrogen bonds are formed or not.

The relative position of the hydrophilic and hydrophobic side is the outcome of two factors: the strength of hydrogen bonds, and the hydrophobicity of the acetate group. If the hydrophilic side is facing the ice surface (figure 7), hydrogen bonds are created between the hydroxyl groups and the ice surface. Conversely if the hydrophobic side is facing the ice, a more stable situation, hydrogen bonds can also be created between the acetate groups and the ice surface. If the acetate anion was isolated, the dipole would be stronger in acetate than in the hydroxyl group, and thus the hydrogen bonds would be stronger for the acetate groups. However, in the polymer structure presented here, the hydroxyl group is probably more polarized than the acetate, and thus we expect the hydrogen bonding to ice to be more favorable with the hydroxyl group.

The second possible mechanism is a configuration similar to the binding of antifreeze proteins to ice, where the hydrophobic side is facing the ice and the hydrophobic acetate groups order water which are in turn anchored to the polymer structure by hydrogen bonds4. It is nevertheless questionable whether such anchored clathrate binding mechanism is compatible with the structure of zirconium acetate. It would be an extraordinary coincidence that the binding mechanism of ZRA and ISPs are the same, considering how different these compounds are.

As opposed to ISPs, the hydroxy-bridged structure of ZRA exhibits very limited flexibility. The spacing between the hydroxyl groups should thus closely match that of the oxygen at the surface of the ice crystals. A higher concentration is thus required for a similar efficiency. If the anchored clathrate mechanism is acting, the acetate hydrophobic groups must organize the water in a structure that provides an excellent three-dimensional match to the ice surface. Depending on the quality of the match, the required concentration for a similar efficiency may vary.

Because it would prevent ZRA from approaching the ice surface, it is essential for this mechanisms that the ice surface does not exhibit any surface charge. Like metal oxides, ice exhibits an isoelectric point at a specific pH, and its surface is charged out of this range. Compounds like ZRA, susceptible to surface interactions, should thus exist in a suitable structure or conformation close the ice isoelectric point.

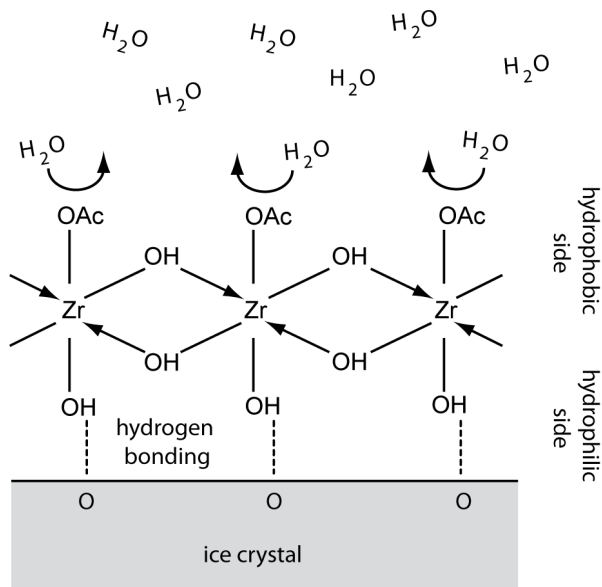

Figure 7. Proposed ice-structuring mechanisms for the hydroxy-bridged polymer structure of zirconium acetate. Hydroxyl groups can be anchored to the surface of ice crystals via hydrogen bonding, leaving the hydrophobic acetate group exposed to water. The binding effectively slows down the incorporation of water molecules onto the crystal surface.

Despite these similarities, the nature of the compounds is fundamentally different: ZRA is a metallic salt, while ISPs are almost systematically long biomacromolecules with a large molecular weight. This opens interesting perspectives for ZRA: its role in ice nucleation, recrystallization and the possible existence of thermal hysteresis, which are all characteristics of antifreeze proteins, has not been investigated so far. The anchoring of zirconium acetate at the ice surface through hydrogen bonds, combined with the non-colligative effects demonstrated previously, may induce a thermal hysteresis during freeze-thaw cycles which should be verified experimentally.

## Conclusions

We propose an ice-structuring mechanism for zirconium acetate which can explain its ice-structuring properties similar to those of ice-structuring proteins. The proposed mechanism is the anchoring of the hydroxy-bridged polymer structure onto the surface of ice via hydrogen bonding of the hydroxyl group, leaving the hydrophobic acetate groups exposed at the surface. This mechanism should effectively slow down the incorporation of water molecules onto the crystal surface, resulting in a smooth interface and a facetted growth. The influence of zirconium acetate on ice nucleation, recrystallization, and the possible existence of thermal hysteresis remains to be investigated.

**AUTHOR INFORMATION**

Corresponding Author

\* Sylvain Deville, e-mail: sylvain.deville@saint-gobain.com

Present Address

† EI-CESI, Rue Magellan, Labège, France

**Author Contributions**

The manuscript was written through contributions of all authors.

**ACKNOWLEDGMENT**

We acknowledge Jérôme Leloup, Arnaud Villard, Gabriel Llorens and Abishek Tiwari for processing some of the samples.

**ABBREVIATIONS**

ZRA, zirconium acetate; ISPs, Ice structuring proteins; PVA, poly-vinyl alcohol.

**REFERENCES**

(1) Davies, P. L.; Baardsnes, J.; Kuiper, M. J.; Walker, V. K. Structure and function of anti-freeze proteins. Phil Trans Royal Soc London B 2002, 357, 927–35.

(2) Clarke, C. J.; Buckley, S. L.; Lindner, N. Ice structuring proteins - a new name for anti-freeze proteins. Cryo letters 2002, 23, 89–92.

(3) The term "ice-structuring" is misleading, as it implies that such proteins or compounds actually modify the crystallographic structure of the ice, which, to the best of our knowledge, has not been reported and would actually be rather extraordinary. The ice is most certainly in the Ih structure; "ice shaping" would thus be a more proper term.

(4) Garnham, C. P.; Campbell; R. L.; Davies, P. L. Anchored clathrate waters bind anti-freeze proteins to ice. Proc. Natl. Acad. Sci. U.S.A. 2011, 108, 7363–7367.

(5) Crilly, J. F.; Russell, A. B.; Cox, A. R.; Cebula, D. J. Designing multiscale structures for desired properties of ice cream. Ind Eng Chem Res 2008, 47, 6362–6367.

(6) Mallick, K. K. Freeze Casting of Porous Bioactive Glass and Bioceramics. J Am Ceram Soc 2009, 92, S85–S94.

(7) Ahmed, A.; Clowes, R.; Myers, P.; Zhang, H. Hierarchically porous silica monoliths with tuneable morphology, porosity, and mechanical stability. J Mat Chem 2011, 21, 5753.

(8) Estevez, L.; Kelarakis, A.; Gong, Q.; Da'as, E. H.; Giannelis, E. P. Multifunctional graphene/platinum/Nafion hybrids via ice templating. J Am Chem Soc 2011, 133, 6122–5.

(9) Gibson, M. I. Slowing the growth of ice with synthetic macromolecules: beyond anti-freeze(glyco) proteins. Polymer Chem 2010, 1, 1141–1152.

(10) Deville, S. et al. Ice shaping properties, similar to that of antifreeze proteins, of a zirconium acetate complex. PLoS ONE 2011, e26474.

(11) Deville, S. et al. Influence of Particle Size on Ice Nucleation and Growth During the Ice-Templating Process. J Am Ceram Soc 2010, 93, 2507–2510.

(12) Lu, K.; Kessler, C. S.; Davis, R. M. Optimization of a Nanoparticle Suspension for Freeze Casting. J Am Ceram Soc 2006, 89, 2459–2465.

(13) Zou, J.; Zhang, Y.; Li, R. Effect of Suspension State on the Pore Structure of Freeze-Cast Ceramics. Int J Applied Ceram Tech 2011, 8, 482–489.

(14) Deville, S.; Saiz, E.; Tomsia, A. P. Ice-templated porous alumina structures. Acta Mater 2007, 55, 1965–1974.

(15) Waschkies, T.; Oberacker, R.; Hoffmann, M. J. Control of Lamellae Spacing During Freeze Casting of Ceramics Using Double-Side Cooling as a Novel Processing Route. J Am Ce-ram Soc 2009, 92, S79–S84.

(16) Nishihara, H.; Iwamura, S.; Kyotani, T. Synthesis of silica-based porous monoliths with straight nanochannels using an ice-rod nanoarray as a template. J Mat Chem 2008, 18, 3662–3670.

(17) Deville, S.; Saiz, E.; Nalla, R. K.; Tomsia, A. P. Freezing as a path to build complex composites. Science 2006, 311, 515–8.

(18) O'Brien, F. Influence of freezing rate on pore structure in freeze-dried collagen-GAG scaffolds. Biomaterials 2004, 25, 1077–1086.

(19) Gutiérrez, M. C. et al. Poly(vinyl alcohol) Scaffolds with Tailored Morphologies for Drug Delivery and Controlled Release. Adv Func Mater 2007, 17, 3505–3513.

(20) Munch, E.; Saiz, E.; Tomsia, A. P.; Deville, S. Architectural Control of Freeze-Cast Ceramics Through Additives and Templating. J Am Ceram Soc 2009, 92, 1534–1539.

(21) Deville, S. et al. In Situ X-Ray Radiography and Tomography Observations of the Solidification of Aqueous Alumina Particles Suspensions. Part II: Steady State. J Am Ceram Soc 2009, 92, 2497–2503.

(22) Deville, S. et al. In Situ X-Ray Radiography and Tomography Observations of the Solidification of Aqueous Alumina Particle Suspensions-Part I: Initial Instants. J Am Ceram Soc 2009, 92, 2489–2496.

(23) Minkoff, I.; Lux, B. Instability criteria for growth of a hopper crystal related to spiral eutectic morphology. J Cryst Growth 1974, 22, 163–165.

(24) Sunagawa, I. Crystals - Growth, Morphology, Perfection. (Cambridge University Press: Cambridge, 2005).

(25) Launey, M. E. et al. Designing highly toughened hybrid composites through nature-inspired hierarchical complexity. Acta Mater 2009, 57, 2919–2932.

(26) Inada, T.; Modak, P. Growth control of ice crystals by poly(vinyl alcohol) and antifreeze protein in ice slurries. Chem Eng Sci 2006, 61, 3149–3158.

(27) Budke, C.; Koop, T. Ice recrystallization inhibition and molecular recognition of ice faces by poly(vinyl alcohol). Chemphyschem 2006, 7, 2601–6.

(28) Wowk, B. et al. Vitrification enhancement by synthetic ice blocking agents. Cryobiology 2000, 40, 228–36.

(29) Inada, T. Substitutes for Antifreeze Proteins : Potential Applications in Ice Slurry Systems. 14th International conference on the properties of water and steam 2004, Kyoto, 660-669.

(30) Knight, C. A.; Wen, D.; Laursen, R. A. Nonequilibrium antifreeze peptides and the recrystallization of ice. Cryobiology 1995, 32, 23–34.

(31) Gutierrez, M. C.; Jobbágy, M.; Rapún, N.; Ferrer, M. L.; Del Monte, F. A Biocompatible Bottom-Up Route for the Preparation of Hierarchical Biohybrid Materials. Adv Mater 2006, 18, 1137–1140.

(32) Zuo, K. H.; Zeng, Y.-P.; Jiang, D. Effect of polyvinyl alcohol additive on the pore structure and morphology of the freeze-cast hydroxyapatite ceramics. Mater Sci Eng C 2010, 30, 283–287.

(33) Pekor, C. M.; Groth, B.; Nettleship, I. The Effect of Polyvinyl Alcohol on the Microstructure and Permeability of Freeze-Cast Alumina. J Am Ceram Soc 2010, 93, 115–120.

(34) Li, N.; Andorfer, C. A.; Duman, J. G. Enhancement of insect antifreeze protein activity by solutes of low molecular mass. J Exp Biol 1998, 201, 2243–51.

(35) Feeney, R. Antifreeze proteins: Current status and possible food uses. Trends Food Sci Technol 1998, 9, 102–106.

(36) Petzold, G.; Aguilera, J. M. Ice Morphology: Fundamentals and Technological Applications in Foods. Food Biophys 2009, 4, 378–396.

(37) Leinala, E. K.; Davies, P. L.; Jia, Z. Crystal structure of beta-helical antifreeze protein points to a general ice binding model. Structure 2002, 10, 619–27.

(38) Pertaya, N. et al. Fluorescence microscopy evidence for quasi-permanent attachment of antifreeze proteins to ice surfaces. Biophysical J 2007, 92, 3663–73.

(39) Pertaya, N.; Marshall, C. B.; Celik, Y.; Davies, P. L.; Braslavsky, I. Direct visualization of spruce budworm antifreeze protein interacting with ice crystals: basal plane affinity confers hyperactivity. Biophysical J 2008, 95, 333–41.

(40) Knight, C. A.; Wierzbicki, A. Adsorption of Biomolecules to Ice and Their Effects upon Ice Growth. 2. A Discussion of the Basic Mechanism of "Antifreeze" Phenomena. Cryst Growth Des 2001, 1, 439–446.

(41) Sakurada, O.; Konishi, H.; Hashiba, M. In Situ Direct-Casting of Alumina Slurry Stabilized with Zirconium Acetate Using Enzyme Reaction. Key Eng Mater 2001, 206, 381–384.

(42) Gandía, L.; Toranzo, R.; Vicente, M.; Gil, A. Non-aggressive pillaring of clays with zirconium acetate. Comparison with alumina pillared clays. Appl Catal A 1999, 183, 23–33.

(43) Sakurada, O.; Hashiba, M. Depletion stabilization of ceramic suspensions with high solid loading in the presence of zirconium oxy-salts. Stud Surf Sci Catal 2001, 132, 375–379.

(44) Furumi, S.; Uchikoshi, T.; Shirahata, N.; Suzuki, T. S.; Sakka, Y. Aqueous Dispersions of Carbon Nanotubes Stabilized by Zirconium Acetate. J Nanosci Nanotechnol 2009, 9, 662–665.

(45) Mater, H. L. V. Method of waterproofing textiles with zirconyl compounds. (1949).

(46) Forouharshad, M.; Montazer, M.; Moghadam, M. B.; Saligheh, O. Preparation of flame retardant wool using zirconium acetate optimized by CCD. Thermochim Acta 2011, 520, 134–138.

(47) Ryu, R.; Gilbert, R.; Khan, S. A. Influence of cationic additives on the rheological, optical, and printing properties of ink-jet coatings. Tappi J 1999, 128–134.

(48) McAlpine, I. Ammonium zirconium carbonate, an alternative insolubiliser for coating binders. TAPPI Proceedings 165–179 (1982).